\documentstyle{article}
\newcommand{\be}{\begin{equation}}
\newcommand{\ee}{\end{equation}}
\newcommand{\bea}{\begin{eqnarray}}
\newcommand{\eea}{\end{eqnarray}}

\newcommand{\nn}{\nonumber}
\newcommand{\eps}{\epsilon}

\newcommand{\dd}{{\mbox d}}
\newcommand{\al}{\alpha}

\newcommand{\lm}{\lambda}

\newcommand{\Gm}{\Gamma}

\newcommand{\pa}{\partial}                                                     

\begin{document}
\begin{titlepage}

\begin{center}
{{\bf
Four-dimensional integration by parts with
differential renormalization as a method of evaluation of Feynman
diagrams
}\\
\vglue 5pt
\vglue 1.0cm
{ {\large V.A. Smirnov}\footnote{E-mail: smirnov@theory.npi.msu.su } }\\
\baselineskip=14pt
\vspace{2mm}
{\it Nuclear Physics Institute of Moscow State University}\\
{\it Moscow 119899, Russia}
\vglue 0.8cm
{Abstract}}
\end{center}
\vglue 0.3cm
{\rightskip=3pc
 \leftskip=3pc
\noindent
It is shown how strictly four-dimensional integration by parts combined
with differential renormalization and its infrared analogue can
be applied for calculation of Feynman diagrams.
\vglue 0.8cm}
\vglue 5cm
\begin{center}
{\em submitted to Teor. Mat. Fiz.}
\end{center}
\end{titlepage}

{\bf 1.}
Integration by parts (IBP) \cite{IBP}
within dimensional regularization \cite{DIMREG} turned out to be
one of the most powerful methods of calculation of Feynman diagrams.
In spite of the fact that dimensional regularization and renormalization
\cite{BM} are commonly used in practical calculations they
happen to be inconvenient in situations
where chiral and super symmetries are involved.
The purpose of this brief note is to present a strictly four-dimensional
version of this method of calculation. It will be shown how (usual)
four-dimensional IBP supplied with technique of
differential renormalization [4--9]
and its infrared (IR) analogue ($\tilde{R}$-operation)
can be used for calculation of diagrams.

In the next section we shall describe prescriptions of differential
renormalization and $\tilde{R}$-operation for simplest diagrams.
Using an example of the master two-loop diagram, we shall show
in Section 3 how four-dimensional IBP is applied for calculations.
In conclusion possible applications of the proposed method
are applied.

{\bf 2.}
Differential renormalization was first defined in coordinate
space [4--8]. It is
possible to translate its recipes into the momentum space language.
Another possibility is to use
homogeneity properties of Feynman integrals and formulate the
corresponding recipe just in momentum space --- see \cite{Z}
where this prescription was formulated for logarithmically
divergent diagrams with the simplest topology of subdivergences.
(Probably, it is more natural to call this renormalization
homogeneous rather than differential.) In particular, for
the simplest one-loop propagator-type Feynman integral\footnote{For
simplicity, we consider Feynman integrals in Euclidean space.}
one has the
following renormalization prescription \cite{Z}:
\be
RF(q) \equiv
R \int \dd^4 k \frac{1}{k^2 (q-k)^2} =
\int \dd^4 k
\ln k^2/\mu^2 \frac{1}{2} q \frac{\pa}{\pa q} \frac{1}{k^2 (q-k)^2}.
\label{R1L}
\ee
Here $\mu$ is a massive parameter that takes into account
finite arbitrariness of renormalization.
The derivative in $q$ is applied before integration so that the
corresponding integral happens to be convergent.

To calculate (\ref{R1L}) it is worthwhile to  introduce  analytic
regularization \cite{Speer}:
\be
RF(q) = \left. \int \dd^4 k \left(-\frac{\dd}{\dd \lm} \right)
\frac{1}{2} q \frac{\pa}{\pa q} \frac{\mu^{2\lm}}{(k^2)^{1+\lm} (q-k)^2}
\right|_{\lm=0} .
\label{R1La}
\ee
When $\lm\neq 0$, we may use the following order: to calculate the integral,
differentiate in $q$, differentiate in $\lm$ and finally put $\lm=0$.
When calculating the integral one uses the four-dimensional one-loop
formula
\be
\int \dd^4 k \frac{1}{(k^2)^{\lm_1} ((q-k)^2)^{\lm_2}}
= \pi^2 G(\lm_1,\lm_2) \frac{1}{(q^2)^{\lm_1+\lm_2-2}},
\label{1L}
\ee
where $G$ is the four-dimensional $G$-function
\be
G(\lm_1,\lm_2) =
\frac{\Gm(\lm_1+\lm_2-2)}{\Gm(\lm_1) \Gm(\lm_2)}
\frac{\Gm(2-\lm_1) \Gm(2-\lm_2)}{\Gm(4-\lm_1-\lm_2)}.
\label{4G}
\ee
In particular,
\be
G(1,1+\lm) = \frac{1}{\lm (1-\lm)} .
\label{4Ge}
\ee
Finally we have
\be
R \int \dd^4 k \frac{1}{k^2 (q-k)^2}
= \pi^2 \left(1 - \ln q^2/\mu^2 \right) \, .
\label{R1L0}
\ee

We shall need as well an IR analogue of differential renormalization.
For the case of dimensional renormalization, the corresponding IR analogue
was developed in \cite{CS}. This operation $\tilde{R}$ removes IR divergences
(off the mass shell) in a quite
similar way as usual dimensional renormalization removes ultraviolet
(UV) divergences. When we combine it with the
$R$-operation itself we obtain the so-called $R^*$-operation \cite{CS}
($R^*=\tilde{R} R$)
which is successfully applied in renormalization group calculations
and in asymptotic expansions of Feynman integrals in various limits
of momenta and masses (see. e.g., \cite{Smi} for a review).

Differential $\tilde{R}$-operation can be defined in momentum space in quite
the same way as differential renormalization in coordinate
space.\footnote{To be more precise, renormalization prescriptions in
coordinate space formulated in refs.~\cite{SZ,S2},
are reformulated as `IR-renormalization' prescriptions in momentum
space, using UV--IR analogy (as it was done in the case of
dimensional $\tilde{R}$-operation --- see \cite{CS,Smi}).}
For the simplest IR-divergent subgraph consisting of two
subsequent massless lines that generate a nonintegrable factor
$1/k^4$, we have the following prescription:
\be
\tilde{R} \frac{1}{k^4}
= \frac{1}{2} \frac{\pa}{\pa k} k \left(
\frac{\ln k^2 /\tilde{\mu}^2}{k^4} \right),
\label{1lir}
\ee
where $\tilde{\mu}$ --- is an IR-renormalization parameter.
Therefore, for the one-loop IR-divergent Feynman integral
\[ \int \dd^4 k \frac{1}{k^4 (q-k)^2} \]
we obtain the following `$\tilde{R}$-normalized' expression:
\be
\tilde{R} \int \dd^4 k \frac{1}{k^4 (q-k)^2}
= \int \dd^4 k
\left( \frac{1}{2} \frac{\pa}{\pa k} k
\frac{\ln k^2 /\tilde{\mu}^2}{k^4} \right) \frac{1}{(q-k)^2} .
\label{RT1L}
\ee
The derivative involved in (\ref{1lir}) and (\ref{RT1L})
is understood in the distributional sense
(that is formally equivalent to the four-dimensional IBP). i.e.
\be
\tilde{R} \int \dd^4 k \frac{1}{k^4 (q-k)^2}
= \int \dd^4 k
\frac{k(k-q) \ln k^2 /\tilde{\mu}^2}{k^4 (q-k)^4} .
\label{RT1La}
\ee

To calculate (\ref{RT1L}) it is helpful to introduce analytic
regularization and then differentiate the integrand at $\lm\neq 0$ using
$\frac{1}{2}\frac{\pa}{\pa k} k \frac{\tilde{\mu}^{2\lm}}{(k^2)^{2+\lm}}
= -\lm \frac{\tilde{\mu}^{2\lm}}{(k^2)^{2+\lm}}.$ We have
\be
\tilde{R} \int \dd^4 k \frac{1}{k^4 (q-k)^2}
= \pi^2 \frac{1}{q^2}
\left(1+ \ln q^2/\tilde{\mu}^2 \right) .
\label{RTln0}
\ee

We shall need also similar integrals with additional logarithms which
are as well calculated with the help of analytic regularization
and one-loop integration formula (\ref{1L}).
One gets
\bea
\tilde{R} \int \dd^4 k \frac{\ln ((q-k)^2/\mu^2)}{k^4 (q-k)^2}
= \int \dd^4 k
\left( \frac{1}{2} \frac{\pa}{\pa k} k
\frac{\ln (k^2 / \tilde{\mu}^2)}{k^4} \right)
\frac{\ln (q-k)^2/\mu^2}{(q-k)^2} \nn \\
= \pi^2 \frac{1}{q^2}
\ln q^2/\mu^2 (\ln q^2/\tilde{\mu}^2 +1) ;
\label{RTln1}
\eea
\bea
\tilde{R} \int \dd^4 k \frac{\ln k^2/\tilde{\mu}^2}{k^4 (q-k)^2}
= \int \dd^4 k
\left( \frac{1}{4} \frac{\pa}{\pa k} k
\frac{\ln^2 k^2 / \tilde{\mu}^2}{k^4} \right)
\frac{\ln (q-k)^2/\mu^2}{(q-k)^2} \nn \\
= \pi^2 \frac{1}{q^2}
\left(1+ \ln q^2/\tilde{\mu}^2 + \frac{1}{2} \ln^2 q^2/\tilde{\mu}^2
\right) ;
\label{RTln11}
\eea
\bea
\tilde{R} \int \dd^4 k \frac{\ln^2 ((q-k)^2/\mu^2)}{k^4 (q-k)^2}
= \int \dd^4 k
\left( \frac{1}{2} \frac{\pa}{\pa k} k
\frac{\ln k^2 / \tilde{\mu}^2}{k^4} \right)
\frac{\ln^2 (q-k)^2/\mu^2}{(q-k)^2} \nn \\
= \pi^2 \frac{1}{q^2} \left( \ln^2 q^2/\mu^2
(\ln q^2/\tilde{\mu}^2 +1) + 4 \zeta(3) \right);
\label{RTln12}
\eea
\bea
\tilde{R} \int \dd^4 k
\frac{\ln ((q-k)^2/\mu^2) \ln k^2/\tilde{\mu}^2}{k^4 (q-k)^2}
= \int \dd^4 k
\left( \frac{1}{4} \frac{\pa}{\pa k} k
\frac{\ln^2 k^2/\tilde{\mu}^2 }{k^4} \right)
\frac{\ln (q-k)^2/\mu^2}{(q-k)^2} \nn \\
= \pi^2 \frac{1}{q^2} \left(
\ln q^2/\mu^2 \left( 1 + \ln q^2/\tilde{\mu}^2
+ \frac{1}{2} \ln^2 q^2/\tilde{\mu}^2\right) + 2 \zeta(3) \right) .
\label{RTln21}
\eea
Here $\tilde{\mu}$ is an IR renormalization parameter
and $\zeta(z)$ Riemann zeta-function.

{\bf 3.}
To calculate the two-loop master massless diagram (see Fig.~1)
strictly in four dimensions
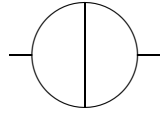
\begin{figure}[htb]
\setlength {\unitlength}{1mm}
\begin{picture}(130,30)(-10,0)
\put (65,15) {\circle{14}}
\put (65,8) {\line(0,1){14}}
\put (55,15) {\line(1,0){3}}
\put (72,15) {\line(1,0){3}}
\end{picture}
\caption{Two-loop master diagram.}
\end{figure}
\be
J(q) = \int\int
\frac{\dd^4 k \dd^4 l}{k^2 (k-q)^2 (k-l)^2 l^2 (l-q)^2}
\label{MD}
\ee
let us apply four-dimensional IBP
\be
\int\int \dd^4 k \dd^4 l \frac{\pa}{\pa l_{\al}} \left\{
\frac{(k-l)_{\al}}{k^2 (k-q)^2 (k-l)^2 l^2 (l-q)^2}
\right\} =0.
\label{BI}
\ee
Formally, this is the same identity that was used within
the standard IBP method \cite{IBP} based on dimensional regularization.
Using relations $2(k-l)l = k^2-(k-l)^2-l^2$,
$2(k-l)(l-q) = (k-q)^2-(k-l)^2-(l-q)^2$, one gets
\be
\int\int \dd^4 k \dd^4 l
\frac{(k-l)^2 - (k-q)^2 }{k^2 (k-q)^2 (k-l)^2 l^2 (l-q)^4} =0.
\label{I1}
\ee
Although this integral is UV and IR convergent one obtains
divergences of both kind when considering separately two terms in the
numerator. It is possible however to separate them by introducing (in
advance) $R$ and $\tilde{R}$ operations.
For example, for the $R$-operation one uses the following
identities (with $j=1,2,\ldots$):
\be \;\;\;\;
\int \dd k \ln^j l^2/ \tilde{\mu}^2 \Pi (k,l,q) =
\int \dd k \ln k^2/\mu^2 \ln^j l^2 / \tilde{\mu}^2
\frac{1}{2} \left(q \frac{\pa}{\pa q}  + l \frac{\pa}{\pa l}
-\lm -4 \right) \Pi (k,l,q) ,
\label{I2}
\ee
where $\Pi$ is a homogenous function of $k,l,q$ of degree $\lm$.
Then one obtains
\be
R^* \int\int \dd^4 k \dd^4 l
\frac{1}{k^2 (k-q)^2 l^2 (l-q)^4}
= R^* \int\int \dd^4 k \dd^4 l
\frac{1}{k^2 (k-l)^2 l^2 (l-q)^4}
\label{I3}
\ee
where $R^*=R \tilde{R}$.

Let us now apply four-dimensional IBP with additional logarithm:
\be
\int\int \dd^4 k \dd^4 l \frac{\pa}{\pa l_{\al}} \left\{
\frac{(k-l)_{\al} \ln l^2/ \tilde{\mu}^2  }{k^2 (k-q)^2 (k-l)^2 l^2 (l-q)^2}
\right\} =0.
\label{BIL}
\ee
After that one comes to
\bea
J(q) =
R \int \dd^4 k \frac{1}{k^2 (q-k)^2}
\tilde{R} \int \dd^4 l \frac{\ln ((q-l)^2/\tilde{\mu}^2)}{l^2 (q-l)^4}
\nn \\
+ R \int \dd^4 k
\frac{1}{k^2 (q-k)^2}
\tilde{R} \int \dd^4 l \frac{\ln l^2/\tilde{\mu}^2}{l^2 (q-l)^4}
-  R^* \int \int \dd^4 k \dd^4 l
\frac{ \ln l^2/\tilde{\mu}^2}{k^2 (k-l)^2  l^2 (q-l)^4}
\nn \\
-  R^* \int \int \dd^4 k \dd^4 l
\frac{ \ln ((l-q)^2/\tilde{\mu}^2)}{k^2 (k-l)^2  l^2 (q-l)^4} .
\label{ID}
\eea
For evaluation of recursively one-loop Feynman integrals involved,
one uses (\ref{R1L0},\ref{RTln0}--\ref{RTln21})
and arrives at the well-known result \cite{Rosner}
\be
J(q) =  6\zeta(3)\pi^4 /q^2 \, .
\label{6z}
\ee

{\bf 4.}
As in the above example of the master two-loop diagram one can apply
four-dimensional IBP with differential $R$ and $\tilde{R}$-operations
at least in every situation where IBP within dimensional
regularization is applied. To do this one uses almost
the same identities as in
the case of dimensional regularization. Roughly speaking, negative powers
of $\eps =(4-d)/2$ (where $d$ is the space-time dimension) are replaced
by some powers of logarithms of the loop momenta. To be able to consider
separately individual terms in these identities one removes artificial
UV and IR divergences in advance (in the case of dimensional regularization
they were automatically regularized). To perform further calculations
one needs a table of one-loop integrals with logarithms (in the
considered case, these are formulae (\ref{R1L0},\ref{RTln0}--\ref{RTln21})).

Note that within analytic regularization  one fails to use
IBP for calculation of the master diagram, as it is possible
within dimensional regularization and, strictly in four dimensions,
in the presented technique. This can be explained by impossibility
of cancellations of the type $p^2/p^2=1$ in the integrands
(because  one obtains $p^2$ in denominators in a power
dependent on the regularization parameter). Therefore, although the table
of the one-loop integrals was derived with the help of
analytic regularization,
the logarithms involved do not spoil these  cancellations
(graphically, corresponding to contraction of lines to a point)
which lead to recursively one-loop integrals in the right-hand side
of (\ref{ID}).

This fact looks promising for construction of differential
renormalization schemes compatible with the gauge invariance because
Ward identities are finally connected with the possibility of the above
cancellations in renormalized Feynman integrals.
It turns out that it is possible to modify the prescriptions
of differential renormalization of refs.~\cite{SZ,S2} in such a way
that the gauge invariance will be preserved automatically (i.e.
without adjusting finite counterterms) at least in the
Abelian case.\footnote{See also ref.~\cite{ZKM}
where relations relevant to Ward identities in QED were proved
in the framework of differential renormalization
for some special classes of diagrams.}
This assertion will be justified in a separate paper.

Finally, let us stress that for applications in theories with chiral and super
symmetries, the four-dimensional technique looks more preferable.

This work is supported by the Russian Foundation for
Basic Research, project 96--01--00654.

{\em Acknowledgments.}
I am grateful to K.G.~Chetyrkin and G.A.~Kravtsova for valuable discussions.

\end{document}